\documentclass{sigchi-ext}
\usepackage[T1]{fontenc}
\usepackage{textcomp}
\usepackage[scaled=.92]{helvet} 
\usepackage{graphicx} 
\usepackage{balance}  
\usepackage{booktabs} 
\usepackage{ccicons}  
\usepackage{ragged2e} 
\usepackage{float}
\usepackage[none]{hyphenat}
\usepackage{marginnote} 

\def\plainkeywords{compliance; habits; long-term; physical activity trackers; fitbit; informatics.}

\title{Long-term Compliance Habits: What Early Data Tells Us}

\numberofauthors{6}
\author{%
  \alignauthor{%
    \textbf{Louis Faust}\\
    \textbf{Priscilla Jim\'enez}\\ 
    \textbf{Aaron Striegel}\\ 
    \textbf{Nitesh V. Chawla}\\ 
    \affaddr{University of Notre Dame}\\
    \affaddr{\textit{Department of Computer Science \& Engineering}}\\
    \affaddr{Notre Dame, Indiana 46556}\\
    \email{lfaust@nd.edu\\
            pjimenez@nd.edu\\
            striegel@nd.edu\\
            nchawla@nd.edu}\\
    }\alignauthor{%
    \textbf{David Hachen}\\ 
    \textbf{Omar Lizardo}\\ 
    \affaddr{University of Notre Dame}\\
    \affaddr{\textit{Department of Sociology}}\\
    \affaddr{Notre Dame, Indiana 46556}\\
    \email{dhachen@nd.edu \\
    olizardo@nd.edu} \\
    }} 

\begin{document}


\maketitle

\RaggedRight{} 

\begin{abstract}
The rise in popularity of physical activity trackers provides extensive opportunities for research on personal health, however, barriers such as compliance attrition can lead to substantial losses in data. As such, insights into student's compliance habits could support researcher's decisions when designing long-term studies. In this paper, we examined 392 students on a college campus currently two and a half years into an ongoing study. We find that compliance data from as early as one month correlated with student's likelihood of dropping out of the study ($p < .001$) and compliance long-term ($p < .001$). The findings in this paper identify long-term compliance habits and the viability of their early detection.
\end{abstract}



\newpage

\section{Introduction}
The growing interest in self-monitoring through physical activity trackers provides an opportunity for personal health to be measured with high granularity and at scale \cite{fawcett2015mining}. The mobility and minimally invasive nature of these wearable devices posit them as ideal tools for long-term monitoring. However, long-term studies are commonly subject to attrition as compliance on behalf of participants can diminish throughout the study. As projects involving daily monitoring of individuals lengthen from months to years to decades, understanding participants compliance habits may assist in the design process of long-term monitoring studies.

In this paper, we explore the correlation between early and long-term compliance behavior. We first identify the appropriate amount of data necessary to separate students based on their compliance and then with this time-frame, we examine its relationship to dropout and long-term habits. Finally, we conclude with the discussion of how knowledge of these relationships may support researchers decisions in designing long-term monitoring studies and our next steps for predicting long-term compliance.

\section{Methods}

\begin{margintable}[5pc]
  \begin{minipage}{\marginparwidth}
    \centering
    \begin{tabular}{r c}
    \multicolumn{1}{r}{Demographic} & $n$ = 392 \\ \hline
    \multicolumn{1}{l|}{male} & 207 (52\%) \\
    \multicolumn{1}{l|}{female} & 185 (47\%) \\ \hline
    \multicolumn{1}{l|}{white} & 261 (66\%) \\
    \multicolumn{1}{l|}{latino} & 51 (13\%) \\
    \multicolumn{1}{l|}{asian} & 36 (9\%) \\
    \multicolumn{1}{l|}{black} & 22 (5\%) \\
    \multicolumn{1}{l|}{foreign} & 22 (5\%)
    \end{tabular}
    \caption{Demographic overview of students. Percentages are in respect to each demographic category.}~\label{demo}
  \end{minipage}
\end{margintable}

\subsection{Participants}
Participants include 392 individuals who entered the university as first-year students in the Fall of 2015. Students' ages ranged from 17 to 19 years. A demographic overview of the students is provided in Table \ref{demo}. 

\subsection{Data collection}
The data used in this paper comes from the NetHealth study conducted at the University of Notre Dame \cite{purta2016}. The study includes an ongoing collection of demographic, psychometric, social network and physical activity data. 
All students were issued a Fibit Charge HR and asked to wear it as much as possible. In addition to monitoring personal health, fitbits can also monitor students compliance, or how much they wore the fitbit. Because the fitbit records the users heart rate every minute, a students daily compliance percentage can be calculated through the sum of minutes in a day a heart rate was detected, divided by the total number of minutes in a day (1440). 

\subsection{Analysis - Defining early compliance}
\begin{figure}[H]
  \centering
  \includegraphics[width=\linewidth]{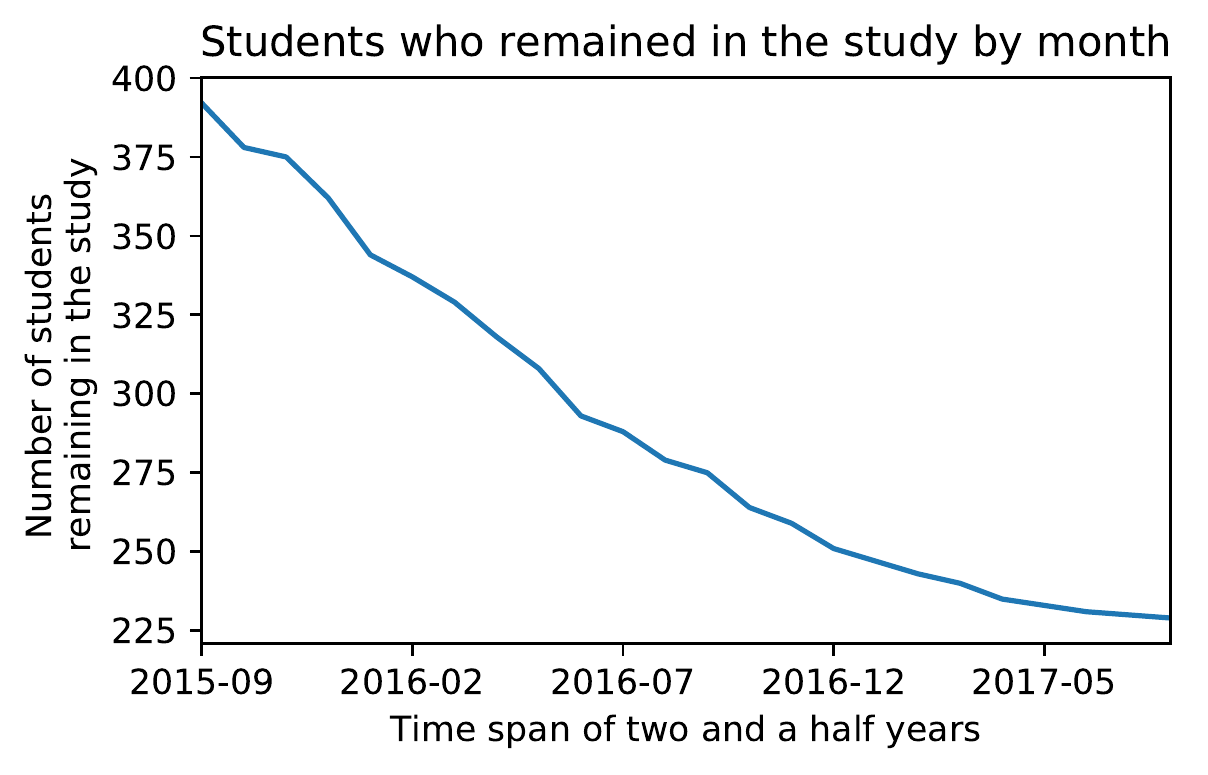}
  \caption{Number of students who remained in the study across two and a half years.}
  \label{dropMonths}
\end{figure}

Of the 392 students participating in the NetHealth study, 166 students have since dropped. Figure \ref{dropMonths} shows that by the sixth month (2016-02) almost 50 students had dropped the study. Since we wanted to focus on students long-term compliance behaviors, we removed dropouts from analysis to prevent a bias as the model may trivially separate students with no compliance data in the later months. However, we return to dropped students in the next section.

\begin{figure}
  \centering
  \includegraphics[width=\linewidth]{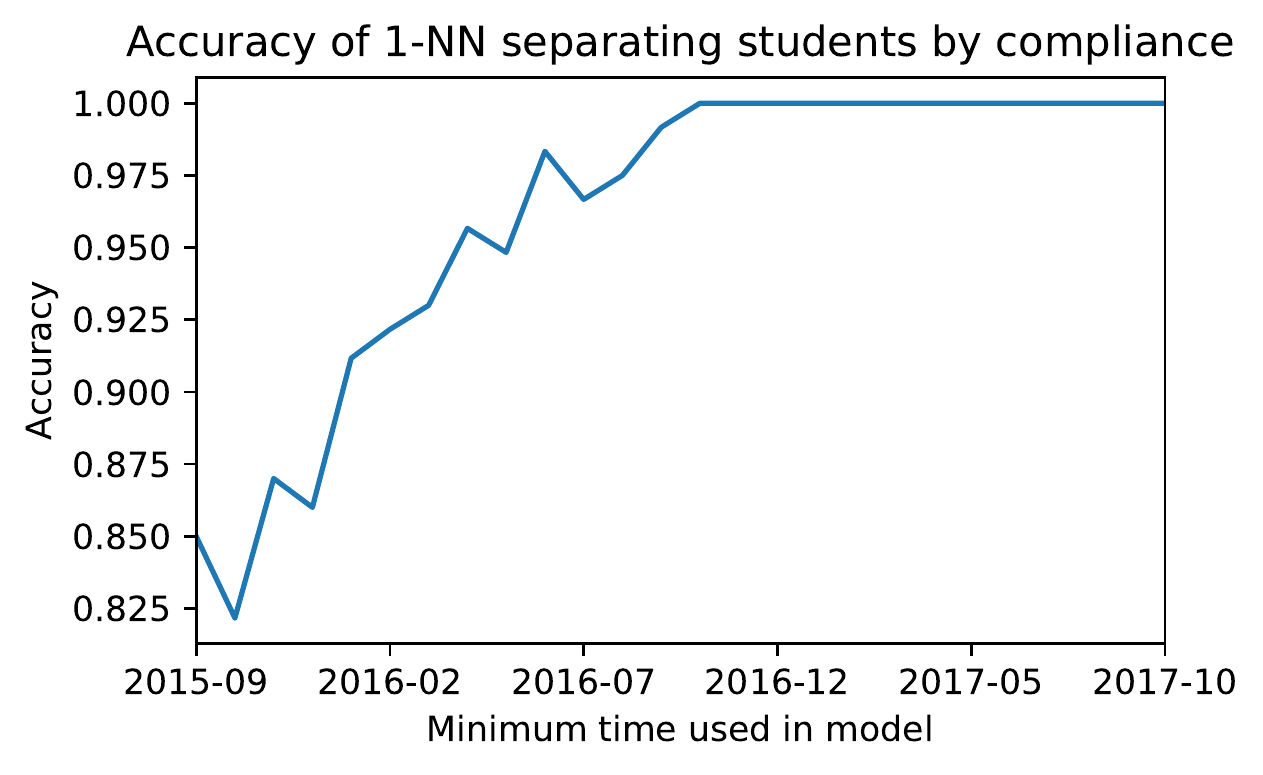}
  \caption{Mean performance of 1-Nearest Neighbor model separating students by compliance as more data is added into the model.}
  \label{KNN}
\end{figure}
 
\newpage 
The remaining 226 students were separated into quartiles based on their total compliance. Total compliance is the sum of days students met the daily compliance threshold throughout their two and a half years in the study. This threshold required a student to wear their fitbit 80\% of the day (19 out of 24 hours) as this provides a good indication of activity and sleep \cite{purta2016}. 

\begin{margintable}[5pc]
  \begin{minipage}{\marginparwidth}
    \centering
    \begin{tabular}{lcc}
    \multicolumn{1}{c}{Dropped?} & \multicolumn{1}{l}{\textbf{Yes}} & \multicolumn{1}{l}{\textbf{No}} \\ \hline
    \multicolumn{1}{l|}{Upper quartile} & 41 & 69 \\
    \multicolumn{1}{l|}{Lower quartile} & 57 & 30
    \end{tabular}
    \caption{Contingency table showing number of students who dropped, separated by compliance during their first month~\label{drop}}
  \end{minipage}
\end{margintable}

To determine the optimal number of months necessary to separate (or predict) students by overall compliance, we employed a 1-Nearest Neighbor (1NN) classifier \cite{scikit-learn}. This method was chosen as the literature shows this to be a robust approach for time series classification \cite{mitsa2010temporal}. The upper and lower quartiles formed the classes for the 1NN model to separate, with 56 students per class. The model was evaluated based on the first month of students compliance (number of days students met the compliance threshold in the first month) using 10-fold cross validation across 100 runs. Consecutive trials were run with each iteration adding a new month of data to the model. Figure \ref{KNN} shows the model's mean accuracy as more data is entered in. We find the model achieved fairly high accuracy with only the first month of data and that by the thirteenth month the classifier reached 100\% accuracy.

Given the performance of the model on only one month of data, we decided to use this as our definition for early compliance. Specifically, we were interested in whether there was a correlation between a students first month of compliance and their long-term compliance behavior and/or likelihood of dropping out. Our results include two separate analyses for exploring these relationships: one including dropped students and the other excluding them.

\section{Results}

\subsection{Dropout}
Including all 392 students in our first analysis, we again separated students by their compliance scores, however, this time it was limited to their first full month in the study: September 2015. Comparing the upper and lower quartiles, 98 students and 99 students respectively, with students who dropped, we found that a significantly higher number of students with poor compliance in their first month ended up dropping the study compared to the students with high compliance (Fisher's Exact, $p < .001$), shown in Table \ref{drop}.

\subsection{Long-term compliance}
\begin{figure*}
  \centering
  \includegraphics[width=\linewidth]{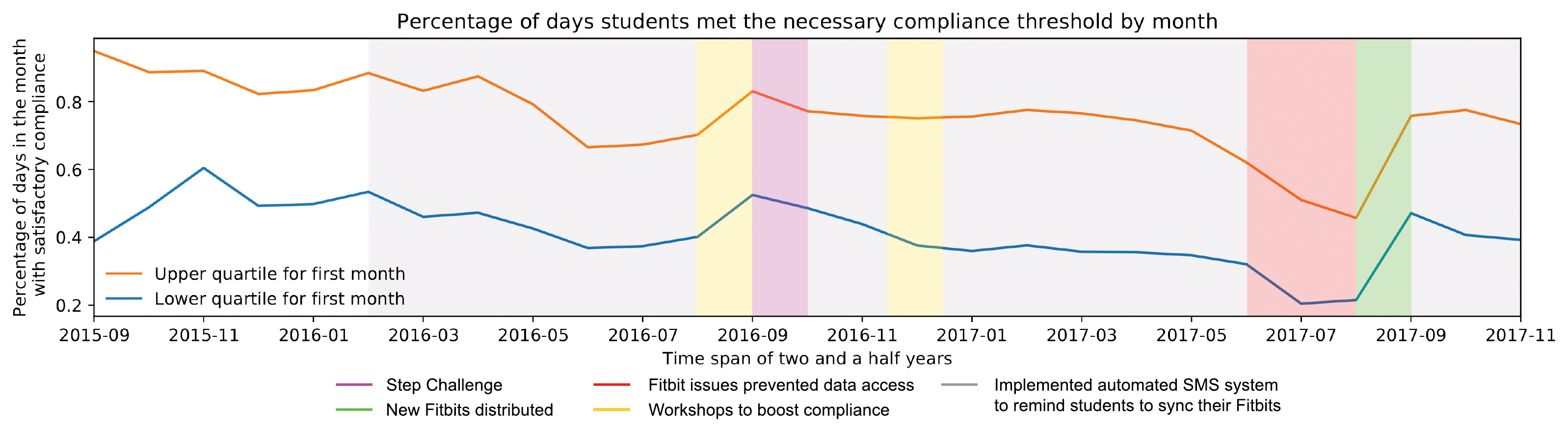}
  \caption{Average compliance across two and a half years. Groups were based on students first month of compliance data. We also highlight interventions conducted to boost compliance and issues encountered to provide context for major shifts in compliance.}
  \label{complyMonths}
\end{figure*}

Given the high number of students who dropped the study, we performed our second analysis using only students who remained in the study for the full two and a half years. Similar to the 1NN analysis, this left 226 students. Again, students compliance rates during the first month of the study were used as the separating variable, forming groups from the upper and lower quartiles with 56 students each.

A visualization of the average compliance rates among these groups is shown over the following 27 months in Figure \ref{complyMonths}. A series of $t-$tests were used to compare the two groups. Results showed students with high compliance in the first month of the study had significantly higher compliance rates per month ($\mu = 34\%, \sigma = 4.5\%$) throughout the next two and a half years compared to students with poor compliance in the first month ($t-$tests, $p < .001$).

\section{Conclusion}
In this paper, we find that compliance habits, good and bad, can persist long-term and be identified within as early as a month. These findings may allow researchers to decide early on in their studies whether to keep students showing poor compliance, or aid them in detecting a group which may require closer monitoring and potential interventions to maintain their compliance and prevent data loss.

We find these habits also show prediction as a viable option when determining students long-term compliance. While we intentionally limited our analysis strictly to students compliance data to see just how much early compliance could tell us about long-term compliance, other factors such as personality have already been linked to compliance and may prove useful as features for a prediction algorithm \cite{faust2017exploring}. We also note that we only considered the most/least compliant students and further study will need to address those remaining students with more vague levels of compliance. We will take these findings into consideration for our next steps towards developing a robust model for forecasting compliance.

\section{Acknowledgements}
The research reported in this paper was supported by the National Heart, Lung, and Blood Institute (NHLBI) of the National Institutes of Health under award number \\ R01HL117757. The content is solely the responsibility of the authors and does not necessarily represent the official views of the National Institutes of Health. 

\newpage

\balance{} 

\bibliographystyle{SIGCHI-Reference-Format}
\bibliography{main}

\end{document}